\documentclass[12pt]{article}
\usepackage{amsmath,amssymb}
\newcommand{\be}{\begin{equation}}
\newcommand{\ee}{\end{equation}}
\newcommand{\bea}{\begin{eqnarray}}
\newcommand{\eea}{\end{eqnarray}}

\begin{document}
\begin{titlepage}
\begin{flushright}
{\,}
\end{flushright}
\vskip 0.9truecm

\begin{center}
{\Large\bf Geometry and Harmonic Superspace: \\ some recent progress} \vspace{1.5cm}

{\large\bf F. Delduc$\,{}^a$, E. Ivanov$\,{}^b$,}\\
\vspace{1cm}

{\it a)Laboratoire de Physique, CNRS et Universit\'e de Lyon, ENS Lyon,}\\
{\it 46, all\'ee d'Italie, 69364 LYON Cedex 07, France}\\
{\tt francois.delduc@ens-lyon.fr}
\vspace{0.3cm}

{\it b)Bogoliubov  Laboratory of Theoretical Physics, JINR,}\\
{\it 141980 Dubna, Moscow region, Russia} \\
{\tt eivanov@theor.jinr.ru}\\

\end{center}
\vspace{0.2cm}
\vskip 0.6truecm  \nopagebreak

\begin{abstract}
\noindent This contribution follows the talk, given by F. Delduc at
the conference SQS'2011 in Dubna, Russia (July 18-23, 2011). To a considerable extent it is a summary of
known facts about the links between geometry and extended
supersymmetry in $d=1$ mechanics, with emphasis on the harmonic
superspace method created in Dubna in the 80's. Some recent
developments based on ref. \cite{DI2011} are also presented.
\end{abstract}
\vspace{0.7cm}

\newpage

\end{titlepage}

\noindent{\large\bf Outline}

\vspace{0.5cm}
\noindent This paper is organized as follows:

\vspace{0.25cm}
\noindent - {\sl\bf Motion of a particle on a manifold :}
An action such that the trajectories are geodesics on some Riemannian manifold is recalled.
\vspace{0.2cm}

\noindent - {\sl\bf Supersymmetric extension and torsion :}
An extension of this model to ${\cal N}=1$ worldline supersymmetry is introduced. The geometry then naturally contains a torsion.
\vspace{0.2cm}

\noindent - {\sl\bf ${\cal N}=2$ supersymmetry :}
Constraints on the geometry such that an extended ${\cal N}=2$ supersymmetry emerges are recalled.
The general solution to these constraints is given. The objects which parametrize this solution, the prepotentials, are used
to construct a superfield action with explicit ${\cal N}=2$ supersymmetry.
\vspace{0.2cm}

\noindent - {\sl\bf ${\cal N}=2$ supersymmetry, particular cases via reduction from $d=2, 4$ :}
The special cases corresponding to K\"ahler and K\"ahler with torsion geometries, which may be obtained by dimensional
reduction from higher dimensions, are recalled.
\vspace{0.2cm}

\noindent - {\sl\bf ${\cal N}=4$ supersymmetry, constraints :}
Geometrical constraints on the geometry such that an extended ${\cal N}=4$ supersymmetry is present are recalled.
\vspace{0.2cm}

\noindent - {\sl\bf ${\cal N}=4$ supersymmetry, particular cases via reduction from $d=2, 4$ :}
The special cases corresponding to hyper-K\"ahler (HK) and hyper-K\"ahler with torsion (HKT) geometries,
which may be obtained by dimensional reduction from higher dimensions, are recalled.
\vspace{0.2cm}

\noindent - {\sl\bf Harmonic superspace :}
Basic facts about harmonic superspace are recalled, with the eventual aim to write down the prepotentials of KT and HKT geometries.
One has to use $2n$ charge-one superfields which describe $n$ hypermultiplets with the off-shell content $({\bf 4,4,0})$.
\vspace{0.2cm}

\noindent - {\sl\bf Superfield constraints and action :}
A general set of superspace constraints and a general superspace action is proposed for $n$ hypermultiplets in $d=1$ mechanics,
following ref. \cite{DI2011}.
\vspace{0.2cm}

\noindent - {\sl\bf Components, bridges and metric :}
Some basic details of expressing geometrical objects (the metric, in particular) in terms of the initial data
in harmonic superspace are described. A thorough study of these geometrical objects leads to the result that the relevant geometry
is analogous to the HKT geometry, apart from the fact that the torsion is not closed. Such a geometry is called weak HKT.
\vspace{0.2cm}

\noindent - {\sl\bf Beyond weak HKT :}
It is conjectured that one might describe more general geometries through the simultaneous use of two different kinds of hypermultiplets.
A calculation in ${\cal N}=2$ superspace sustaining this conjecture is outlined.

\vspace{0.5cm}
\noindent{\large\bf Motion of a particle in a Riemannian manifold $M$}\\

\vspace{0.25cm}
\noindent We consider a differentiable manifold $M$ and a set of local coordinates $x^i,\, i=1\cdots n\,,$ on $M$.
A particle will follow a trajectory parametrized by coordinates $x^i(t)$ depending on time $t$.
This trajectory may be obtained as a minimum of the action
\be S[x]=\int dtg_{ij}(x)\dot x^i\dot x^j\,,\qquad \dot x^i=\frac{dx^i}{dt}\,,\ee
where $g_{ij}(x)$ are the components of a metric tensor on the manifold $M$. The equations of motion are given by
\be\ddot x^i+\gamma^i_{jk}\dot x^j\dot x^k=0\,,\label{eog}\ee
where $\gamma^i_{jk}(x)$ are the Christoffel symbols associated with the metric $g_{ij}(x)$
\be\gamma^i_{jk}=\frac{1}{2}g^{il}(\partial_jg_{lk}+\partial_kg_{lj}-\partial_lg_{jk})\,.\ee
The equations (\ref{eog}) are the equations of geodesics in a particular parametrization, such that the velocity
vector has a constant length along the trajectory, $\dot x^i\dot x^j g_{ij} = const\,.$

 \vspace{0.5cm}
\noindent{\large\bf Supersymmetric extension and torsion}\\

\vspace{0.25cm}
\noindent We now consider a superspace with coordinates  $(t,\theta)$, where $\theta$ is a real Grassmann variable.
Supersymmetry transformations are realized as particular translations in superspace,
with $\delta\theta= \epsilon\,$, $\delta t=-i\epsilon\theta\,$, and $\epsilon$ is a real Grassmann parameter.
The anticommutator of two supersymmetry transformations is a time translation.
We introduce superfields $X^i(t,\theta)$ such that their first components
$x^i(t)=X^i(t,\theta)\vert_{\theta=0}$ give back the coordinates of the particle at time $t$.
We shall use the supersymmetric derivative :
\be D=\frac{\partial}{\partial\theta}+i\theta\frac{\partial}{\partial t}\,,\qquad D^2=i\frac{\partial}{\partial t}\,.\ee
We then write a general supersymmetric action, constrained by the requirement that bosonic component fields
have a field equation of second order in time derivatives:
\be S[X]=\int dtd\theta(ig_{ij}(X)\dot X^iDX^j+\frac{1}{3!}c_{ijk}(X)DX^iDX^jDX^k)\,,\label{n1act}\ee
where $c_{ijk}(x)$ is an antisymmetric tensor which will play the role of a torsion.
In particular, the field equations involve the following covariant derivatives (written for an arbitrary vector field $V^j$)
\be \nabla_iV^j=\frac{\partial}{\partial x^i}V^j+\Gamma^j_{ik}V^k,\label{covdt}\ee
where the connexion reads
\be\Gamma^j_{ik}=\gamma^j_{ik}+\frac{1}{2}g^{jl}c_{ikl}\,.\label{cont}\ee
It contains, as a symmetric part, the Christoffel symbols previously introduced, and, as an antisymmetric part, the new torsion tensor
$c_{ikl}$. It is still a metric connexion, meaning that the covariant derivative of the metric vanishes.

Thus, given any geometry defined by a metric and a torsion, there is an ${\cal N}=1$ supersymmetric action encoding this geometry.

 \vspace{0.5cm}
\noindent{\large\bf ${\cal N}=2$ supersymmetry}\\

\vspace{0.25cm}
\noindent We now look for conditions on the geometry, such that extended worldline ${\cal N}=2$ supersymmetry is in fact present.
The way to do that may be found in a 1980 article by L. Alvarez-Gaum\'e and D. Freedman \cite{AGF}. We consider a general
form of the transformations under the second supersymmetry, such that it automatically anticommutes with the first supersymmetry
\be\delta X^i=\epsilon' J^i_j(X)DX^j\,,\label{n2tr}\ee
where $\epsilon'$ is an extra Grassmann parameter and $J^i_j(x)$ is a tensor on $M$. There are now two sources of constraints
on the tensor $J$.

The first one comes from requiring that the new transformations form a supersymmetry algebra. This leads
to the equations
\be J^i_jJ^j_k=-\delta^i_k\,,\quad J^l_i\frac{\partial}{\partial x^{[l}}J^k_{j]}-J^l_j\frac{\partial}{\partial x^{[l}}J^k_{i]}=0\,,
\label{ics}\ee
which are summarized by saying that the tensor $J$ is an integrable complex structure.

The second source of constraints comes from requiring that the transformations (\ref{n2tr}) leave invariant the action (\ref{n1act}).
This leads to three equations. The first one is
\be g_{ik}J^k_j+J^k_ig_{kj}=0\,,\ee
and it means that the metric is hermitian with respect to the complex structure. The second equation is
\be\nabla_iJ^k_j+\nabla_jJ^k_i=0\,,\label{scdcs}\ee
and it means that the symmetrized covariant derivatives of the complex structure vanish. The covariant derivatives are just
those introduced in (\ref{covdt}), (\ref{cont}). Finally, the third equation reads
\be \partial_{[i}(J^m_jc_{kl]m})-2J^m_{[i}\partial_{[m}c_{jkl]]}=0\,.\label{4cs}\ee
It tells us that some 4-form, linear in the torsion $c$ and the complex structure $J$, vanishes.
All these results may be found in a 1990 article by R. Coles and G. Papadopoulos \cite{CoPa}.

It turns out that the constraints (\ref{ics})-(\ref{4cs}) may easily be solved. From (\ref{ics}) it follows
that there exist local complex coordinates $(z^\alpha,\bar z^{\bar\alpha})\,$, such that the complex structure is constant
\be J_\alpha^\beta=i\delta_\alpha^\beta\,,\quad J_{\bar\alpha}^{\bar\beta}=-i\delta_{\bar\alpha}^{\bar\beta}\,, \quad
J_\alpha^{\bar\beta}=J_{\bar\alpha}^\beta=0\,,\ee
and the change of coordinates leading from one patch to another has to be holomorphic. In these complex coordinates,
the metric has only mixed components $g_{\alpha\bar\beta}\,$, $g_{\alpha\beta}=g_{\bar\alpha\bar\beta}=0\,$.
Finally, from the remaining two constraints (\ref{scdcs}), (\ref{4cs}) one may show that the torsion is fully specified
in terms of the metric and a 2-form $B_{\alpha\beta}\,$, $\bar B_{\bar\alpha\bar\beta}\,$, with the mixed components vanishing.

One may then write an action for this geometry which has explicit ${\cal N}=2$ supersymmetry. One uses an ${\cal N}=2$ superspace with
coordinates $(t,\theta,\bar\theta)$, where $\theta$ is now a complex Grassmann variable, and supersymmetric derivatives are
\be D=\frac{\partial}{\partial\theta}+i\bar\theta\frac{\partial}{\partial t}\,,\,\,
\bar D=\frac{\partial}{\partial\bar\theta}+i\theta\frac{\partial}{\partial t}\,,\,\,\,\, D^2=\bar D^2=0\,,\,\,
 \{D,\bar D\}=2i\frac{\partial}{\partial t}\,.\ee
The coordinates $z^\alpha$, $\bar z^{\bar\alpha}$ are defined as the first components of ${\cal N}=2$ superfields
$Z^\alpha$, $\bar Z^{\bar\alpha}\,$, which satisfy the chirality constraints $\bar DZ^\alpha=0\,$, $D\bar Z^{\bar\alpha}=0\,$.
The most general action reads
\be S[Z,\bar Z]=\int dtd\theta d\bar\theta(g_{\alpha\bar\beta} DZ^\alpha\bar D\bar Z^{\bar\beta}
+B_{\alpha\beta} DZ^\alpha DZ^{\beta}+\bar B_{\bar\alpha\bar\beta}\bar D\bar Z^{\bar\alpha}\bar D\bar Z^{\bar\beta})\,.\label{n2act}\ee
This action is written in terms of the unconstrained objects    $g_{\alpha\bar\beta}\,$, $B_{\alpha\beta}\,$, $B_{\bar\alpha\bar\beta}\,,$
which determine the geometry. It is a general fact that the actions with explicit extended supersymmetry
are written in terms of the unconstrained data (called prepotentials) which determine the geometry.
This general ${\cal N}=2$ superspace action (\ref{n2act}) may be found, together with many other related results,
in a 1999 article by C. Hull \cite{Hu}.\footnote{The ${\cal N}=2$ supersymmetric quantum mechanics associated with the action (\ref{n2act})
also exhibits interesting geometric properties which were recently analyzed in \cite{ism}.}

\vspace{0.5cm}
\noindent{\large\bf ${\cal N}=2$ supersymmetry, particular cases via reduction from $d=2, \;4$}\\

\vspace{0.25cm}
\noindent Among these ${\cal N}=2$ geometries in supersymmetric mechanics, some special cases originate from theories in dimension
$d=2$ and $d=4$
through the dimensional reduction to $d=1$. We recall that, in two dimensions, one separates right-handed and left-handed
supersymmetries and uses the symbol ${\cal N}=(p,q)$ to denote them.
\begin{itemize}
\item{
${\cal N}=1$ supersymmetry, $d=4$ (or ${\cal N}=(2,2)$ supersymmetry, $d=2$): Torsion vanishes, covariant derivatives
of the complex structure vanish.
It corresponds to the celebrated K\"ahler geometry. In this case, the metric may be written as a second derivative
\be g_{\alpha\bar\beta}=\partial_\alpha\partial_{\bar\beta}K(z,\bar z)\,,\ee
where the scalar function $K(z,\bar z)$ is called the K\"ahler potential. The K\"ahler potential is not necessarily
defined as a function on the whole manifold. It may change, when going from the patch $U_{(a)}$ to another patch $U_{(b)}\,$, as
\be K_{(b)}(z_\beta,\bar z_\beta)=K_{(a)}(z_\alpha,\bar z_\alpha)+f(z_\alpha)+\bar f(\bar z_\alpha)\,.\ee
In $d=4$ and $d=2$, the superspace action is directly determined by the K\"ahler potential. This was the first example
of an action given in terms of the prepotential of the target geometry, and it can be found in
a 1979 paper by B. Zumino \cite{Zum}}.

\item{
${\cal N}=(2,0)$, $d=2$: Torsion is a closed 3-form, covariant derivatives of the complex structure vanish. This geometry is called
K\"ahler with torsion (KT). The prepotential of this geometry has a vector index
\be g_{\alpha\bar\beta}=\partial_{\bar\beta} V_\alpha+\partial_\alpha V_{\bar\beta}\,,\ee
and the torsion tensor is also determined in terms of the vector prepotential  $V_\alpha\,, \,V_{\bar\beta}\,$, which
is not a globally defined vector field, however.
The K\"ahler geometry appears as a special case of the KT geometry, when the vector potential $V$ is expressible as a derivative
of the scalar potential $K\,$,
\be V_\alpha=\frac{1}{2}\frac{\partial}{\partial z^\alpha}K\,, \quad V_{\bar\alpha}=
\frac{1}{2}\frac{\partial}{\partial\bar z^{\bar\alpha}}K\,.\ee}
This geometry was described in a 1985 article by C. Hull and E. Witten \cite{HuWi}.
\end{itemize}

\newpage
\noindent{\large\bf ${\cal N}=4$ supersymmetry, constraints}\\

\vspace{0.25cm}
\noindent The constraints which ensure the action (\ref{n1act}) to possess ${\cal N}=4$ supersymmetry form a natural generalization
of the ${\cal N}=2$ constraints \cite{CoPa}. Since we use a formalism with one supersymmetry being explicit,
we need three additional supersymmetry transformations anticommuting with the first one. We thus introduce 3 tensors $J_a$, $a=1\cdots 3$,
and write the infinitesimal transformations as
\be \delta X^i=\sum_{a=1}^3\epsilon^a(J_a)^i_jDX^j,\label{3es}\ee
where $\epsilon^a$, $a=1,2,3\,$, are three real Grassmann parameters.

Again, a first set of constraints comes from requiring that
the new transformations form the supersymmetry algebra and anticommute with each other. One finds
\be J_aJ_b+J_bJ_a=-2\delta_{ab}\mathbf{1}\,, \quad (J_a)^l_i\frac{\partial}{\partial x^{[l}}(J_b)^k_{j]}
-(J_a)^l_j\frac{\partial}{\partial x^{[l}}(J_b)^k_{i]}+(a\leftrightarrow b)=0\,.\ee
Thus the tensors $J_a$ are three integrable complex structures, which anticommute with each other.

A second set of constraints comes from requiring the invariance of the action (\ref{n1act})
under the transformations  (\ref{3es}). First, the metric has to be hermitian with respect to all three complex structures
\be g_{ik}(J_a)^k_j+(J_a)^k_ig_{kj}=0\,,\ee
second, the symmetrized derivatives of all three complex structures should vanish,
\be D_i(J_a)^k_j+D_j(J_a)^k_i=0\,,\ee
and, finally, three 4-forms made out of the complex structures and the torsion should vanish
\be\partial_{[i}((J_a)^m_jc_{kl]m})-2(J_a)^m_{[i}\partial_{[m}c_{jkl]]}=0\,.\ee

The resolution of these constraints is much more difficult than in the ${\cal N}=2$ case.
Some particular cases are known, and now we shall recall them.

 \vspace{0.5cm}
\noindent{\large\bf ${\cal N}=4$ supersymmetry, particular cases via reduction from $d=2,\; 4$}\\

\vspace{0.25cm}
\noindent Some of the mechanical models with ${\cal N}=4$ supersymmetry may be obtained by dimensional reduction from models
in a higher dimension.
\begin{itemize}
\item{
${\cal N}=2$ supersymmetry, $d=4$ (or ${\cal N}=(4,4)$ supersymmetry, $d=2$): Torsion vanishes, all three complex structures are annihilated
by covariant derivatives and form the quaternionic algebra
\be J_aJ_b=-\delta_{ab}\mathbf{1}+\epsilon_{abc}J_c\,.\ee
This particular geometry is called the Hyper-K\"ahler (HK) geometry.}
\item{
${\cal N}=(4,0)$, $d=2$: Torsion is a closed 3-form, complex structures are annihilated by covariant derivatives
(with a connexion including torsion) and form the quaternionic algebra. This geometry is called the Hyper-K\"ahler with torsion (HKT) geometry.}
\end{itemize}
In both cases, the prepotentials of the geometry are known.
They have been studied by A. Galperin, E. Ivanov, S. Kalitzin, V. Ogievetsky  and E. Sokatchev \cite{hkHSS} for the HK geometry
and by F. Delduc, S. Kalitzin, E. Sokatchev  \cite{dks} for the HKT geometry. Both cases require making use of harmonic superspace.

 \vspace{0.5cm}
\noindent{\large\bf Harmonic superspace \cite{HSS,HSS1}}\\

\vspace{0.25cm}
\noindent The coordinates of ${\cal N}=4$ superspace in one dimension may be written as $(t,\theta^i,\bar\theta_i)$, where $\theta^i$, $i=1,2\,,$
is a pair of complex Grassmann variables. In the harmonic approach, one adds to these variables another set of bosonic variables
$(u^\pm_i),\,i=1,2$, called harmonic variables. They should be such that the 2 by 2 matrix
\be \left(\begin{array}{cc}u^+_1 &u^-_1\cr u^+_2& u^-_2\end{array}\right)\ee
belongs to the group SU(2). All fields depend on harmonic variables, and have definite charges
under the right action of the diagonal U(1) subgroup of SU(2). In harmonic superspace, one can find a subspace
invariant under all four supersymmetries
\be (t_A, \theta^+=\theta^iu^+_i,\bar\theta^+=\bar\theta^iu^+_i,u^\pm_i)\,.\ee
This subspace is called analytic superspace. It involves only half of the original Grassmann variables.
To describe HK or HKT geometry, one needs a set of $2n$ charge $1$ analytic superfields (which will be called hypermultiplets)
\be q^{+a}(t_A,\theta^+,\bar\theta^+,u^\pm),\,\, a=1\cdots 2n\,.\ee
In the HK case, the prepotential is a charge 4 scalar function
${\cal L}^{+4}(q^{+a},u^\pm)\,$.
In the HKT case, the analytic prepotential carries an $Sp(2n)$ index and has charge 3,
$ {\cal L}^{+3a}(q^{+b},u^\pm)$.
Notice that HK is a special case of HKT, with
\be {\cal L}^{+3a}=\Omega^{ab}\frac{\partial}{\partial q^{+b}}{\cal L}^{+4},\ee
where $\Omega$ is a $2n$ by $2n$ constant regular antisymmetric matrix (also called a symplectic metric).
Since in harmonic superspace there are new coordinates, the harmonic variables $u^\pm_i$,
there also appear new derivatives, which are consistent with the constraints on the harmonic variables.
They are called harmonic derivatives and read
\be D^{++}=u^{+i}\frac{\partial}{\partial u^{-i}},\,\,
D^{--}=u^{-i}\frac{\partial}{\partial u^{+i}},\,\,
D^{0}=u^{+i}\frac{\partial}{\partial u^{+i}}-u^{-i}\frac{\partial}{\partial u^{-i}}\,.\ee
Their commutation relations give back the Lie algebra of SU(2).
An important point about the derivative operator $D^{++}$ is that it acts inside the analytic subspace.

\vspace{0.5cm}
\noindent{\large\bf Superfield constraints and action }\\

\vspace{0.25cm}
\noindent In two dimensions, the field equations of an HKT nonlinear sigma model read
\be D^{++}q^{+a}={\cal L}^{+3a}(q^{+},u^\pm)\,. \label{nlc}\ee
When restricted to one dimension, this equation is no longer dynamical. It puts to zero some component fields
inside $q^{+a}$, but it does not restrict the time dependence of those component fields which survive. It
is a harmonic constraint which restrict the SU(2) content of the superfields. It may be shown that the content of
the superfields $q^{+a}$, subject to the constraints (\ref{nlc}) (as well as to some self-consistent reality condition) is just
that of $n$ $({\bf 4,4,0})$ multiplets of $d=1\,$, ${\cal N}=4$ supersymmetry in one dimension. Each $({\bf 4,4,0})$ multiplet
contains $4$ real bosons and $4$ real fermions.

Since the constraints (\ref{nlc}) are not dynamical, we need some extra input from which the equations of motion of the fields
may be obtained. The most general action leading to equations for the physical bosons which are of second order in time derivatives reads
\be S=\int dtd^4\theta du\,{\cal L}(q^{+a},q^{-a}, u^\pm)\,,\quad q^{-a}=D^{--}q^{+a}.\label{ga}\ee
Since this action is integrated over the whole superspace, it is not required that the integrand lives on the analytic subspace.
Indeed, the lagrangian density $\cal L$ depends on the non analytic superfields $q^{-a}$.
One may add to this action a term which is an integral on the analytic subspace only
\be S_{WZ}=\int dudt_Ad^2\theta^{(-2)}\,{\cal L}^{+2}(q^{+a}, u^\pm)\,.\ee
It is called a Wess-Zumino term, and physically it describes the coupling of the particle to an external magnetic field.
The major part of the article \cite{DI2011} is devoted to extracting the geometry experienced by the physical fields
from the non-linear constraints (\ref{nlc}) and the action (\ref{ga}).

\vspace{0.5cm}
\noindent{\large\bf Components, bridges and metric}\\

\vspace{0.25cm}
\noindent One first expands the superfields $q^{+a}$ in powers of the Grassmann variables $\theta^+$, $\bar\theta^+$
\be q^{+a}=f^{+a}(t,u)+\theta^+\chi^a(t,u) +\bar\theta^+\bar\chi^a(t,u)+\theta^+\bar\theta^+ A^{-a}(t,u)\,.\ee
It may be shown that the component $A^{-a}$ is fully determined by the other components as a consequence of the constraint (\ref{nlc}).
The remaining components are not yet ordinary fields. They depend not only on time, but also on harmonic variables $u^\pm_i$.
This dependence is restricted as a consequence of the constraint (\ref{nlc})
\be D^{++}f^{+a}={\cal L}^{+3a}(f^{+},u^\pm)\,,\quad
D^{++}\chi^{a}-\frac{\partial {\cal L}^{+3a}(f^{+},u^\pm)}{\partial f^{+b}}\chi^b=0\,. \label{nlcc}\ee
The first of the equations (\ref{nlcc}) means that $f^{+a}$ is determined if one knows its lowest order term in harmonic variables.
One has to separate
\be f^{+a}(t,u)=f^{ia}(t)u^+_i+v^{+a}(f^{jb}(t),u^\pm)\,.\ee
The $4n$ fields $f^{ia}(t)$ form coordinates of the manifold under study. The functions $v^{+a}$ may be interpreted as
a bridge between two different sets of coordinates.

The fermionic components $\chi^{a}(t,u)$ and its complex conjugate also satisfy a complicated harmonic equation,
which is the second equation of (\ref{nlcc}). It may be simplified by introducing a frame bridge, which is
a $2n\times 2n$ matrix $M$ satisfying
\be D^{++}M^{\underline b}_a+\frac{\partial {\cal L}^{+3c}}{\partial f^{+a}}M^{\underline b}_c =0\,.\ee
Then the fermionic field $\chi^{\underline a}=M^{\underline a}_b\chi^b$ is independent of harmonic variables,
$D^{++}\chi^{\underline a}=0\,$, and thus depends only on time. The frame bridge is used to define the harmonic
independent vielbeins $e^{\underline k\underline a}_{ib}$ as
\be\frac{\partial f^{+a}}{\partial f^{ib}}M^{\underline a}_a=-e^{\underline k\underline a}_{ib}u^+_k\,,\ee
as well as the symplectic metric
\be G_{\underline a\underline b}=\int du(M^{-1})^{a}_{\underline a}
(M^{-1})^{b}_{\underline b}(\partial_{+[a}\partial_{-b]}{\cal L}+\cdots)\,.\ee
One finally gets the local expression  for the Riemannian metric on the manifold
\be g_{ia\, kb}=G_{\underline c\underline d}\epsilon_{\underline l\underline t}
e_{ia}^{\underline l\underline c}e_{kb}^{\underline t\underline d}\,.\ee
Notice that the tangent space metric $G_{\underline c\underline d}$ is not constant, so the vielbeins $e^{\underline k\underline a}_{ib}$
do not define an orthonormal frame. On the contrary, complex structures are constant in the tangent space and read in the coordinate space as
\be (J_{(\underline l\underline k)})^{ia}_{jc}=ie^{ia}_{(\underline l\underline b}e_{jc}^{\underline t\underline b}
\epsilon_{\underline k)\underline t}\,.\ee
Finally, the component action reads
$$S=\int dt\left[\frac{1}{2}\,g_{ia\, kb}\,\dot{f}^{ia}\dot{f}^{kb}
-\frac{i}{4}\,G_{[\underline{a}\,\underline{b}]}\left(\nabla \bar{\chi}^{\underline{a}}\chi^{\underline{b}}
- \bar{\chi}^{\underline{a}}\nabla \chi^{\underline{b}}\right) - \frac{1}{16}
\left(\epsilon^{\underline{i}\,\underline{k}}\nabla_{\underline{i}[\underline{a}}\nabla_{\underline{k}\underline{b}]}
\,G_{[\underline{c}\,\underline{d}]}\right)\bar\chi{}^{\underline{a}}
\bar\chi{}^{\underline{b}}\chi^{\underline{c}}\chi^{\underline{d}}\right].$$
The salient points of the results that were obtained in \cite{DI2011} are that complex structures form
a quaternionic algebra, that they are covariantly constant and that torsion is in general not closed.
A geometry with such properties was called weak HKT  in a paper by P. Howe and G. Papadopoulos in 1996 \cite{hkt}.
A novel feature which is brought in by the harmonic superspace approach is that this weak HKT geometry
is solved in terms of two unconstrained prepotentials, the general one ${\cal L}(f^+,f^-,u^\pm)$ and
the analytic one ${\cal L}^{+ 3a}(f^+, u^\pm)\,$.

Some particular cases may arise. If the lagrangian in (\ref{ga}) is quadratic, ${\cal L}=\Omega_{ab}q^{+a}q^{-b}$,
then the torsion is closed and the geometry is HKT. If, moreover, the analytic prepotential ${\cal L}^{+3a}$ is a derivative,
\be {\cal L}^{+3a}=\Omega^{ab}\frac{\partial}{\partial q^{+b}}{\cal L}^{+4}\,,\label{curl}\ee
then the geometry is HK. If, however, one restricts the analytic prepotential as in (\ref{curl})
but keeps a general lagrangian ${\cal L}$, one gets a geometry intimately connected to the hyperk\"ahler geometry encoded in  ${\cal L}^{+4}$,
but which includes torsion. In particular, if the manifold has dimension 4, the HKT metric is conformal to the HK metric,
with a conformal factor which is a harmonic function on the HK manifold (i.e. satisfies the covariant Laplace-Beltrami equation on this
manifold, which just amounts to the torsion closedness condition in this case) \cite{chs}. If the conformal factor is arbitrary,
one faces a weak HKT geometry.  In the simplest case ${\cal L}^{+4} = 0$ the metric is conformal to the flat $\mathbb{R}^4$ metric,
while the torsion closedness condition is just the $\mathbb{R}^4$ Laplace equation for the conformal factor \cite{Hu,Strom,IL}.

\newpage
\noindent{\large\bf  Beyond weak HKT}\\

\vspace{0.25cm}
\noindent Thus  a set of hypermultiplets of the same kind does not allow to describe in superspace the most general
geometry allowed by ${\cal N}=4$
supersymmetry in one dimension. We conjecture that the description of this general case requires the simultaneous use of two different types
of hypermultiplets. The automorphism group of the ${\cal N}=4$ supersymmetry algebra is SO(4)$\simeq$SU(2)$\times$SU(2).
One of these two  SU(2) groups acts on the harmonic variables. One may define two types of hypermultiplets, depending
on which SU(2) group is associated with harmonic variables. Very probably, when using the two types together,
one  may describe the general ${\cal N}=4$ geometry.

A computation in support of this conjecture was done in ${\cal N}=2$ superspace \cite{DI2011}. Starting from chiral superfields $z^\alpha,\, y^a$
$(\alpha = 1, \ldots 2n\,, \; a = 1, \ldots 2m)\,,$ one can write 2 extra supersymmetry transformations as
$$\delta z^\alpha=\epsilon J^\alpha_\beta Dz^\beta\,,$$
$$\delta y^a=\bar\epsilon \tilde J^a_b Du^b\,.$$
Then the $z$ coordinates and the $y$ coordinates belong to different $({\bf 4, 4, 0})$ representations of ${\cal N}=4$ supersymmetry.
We have checked that, generically, the complex structures (in the full target space of complex dimension $2(n+m)$) do not form
the quaternionic algebra, and only symmetrized covariant derivatives of complex structures vanish. It remains to show that one indeed
can get the most general geometry in this way. For the particular case of two linear $({\bf 4, 4, 0})$ multiplets of different sorts
(thus corresponding to 8-dimensional target space) the most general component action was constructed in \cite{ils}, proceeding from
${\cal N}=4$ superfield formalism. The set of relevant target metrics encompasses some examples which were explicitly given
earlier in \cite{GiPaSt} and were argued in \cite{Hu} to correspond to the general geometry.

\vspace{0.5cm}
\noindent{\large\bf Acknowledgements.} E.I. acknowledges support from a grant of Heisenberg-Landau Programme and RFBR grants
09-01-93107 and 11-02-90445.

\end{document}